\documentstyle[prl,tighten,floats,aps,epsf,epsfig,psfig]{revtex}
\begin{document} \preprint{IC/99/22} \newcommand{\newc}{\newcommand}
\def\be{\begin{equation}} \def\ee{\end{equation}} \def\bea{\begin{eqnarray}}
\def\eea{\end{eqnarray}} \def\simlt{\stackrel{<}{{}_\sim}}
\def\simgt{\stackrel{>}{{}_\sim}} 

\twocolumn[\hsize\textwidth\columnwidth\hsize\csname @twocolumnfalse\endcsname

\title{Weak-Scale Hidden Sector and Energy Transport in Fireball Models of
Gamma-Ray Bursts}

\author{Durmu\c s A.  Demir and Herman J.  Mosquera Cuesta}

\address{The Abdus Salam International Centre for Theoretical Physics, I-34100
Trieste, Italy}

\date{\today}

\maketitle

\begin{abstract} The annihilation of pairs of very weakly interacting particles
in the neibourghood of gamma-ray sources is introduced here as a plausible
mechanism to overcome the baryon load problem.  This way we can explain how
these very high energy gamma-ray bursts can be powered at the onset of very
energetic events like supernovae (collapsars) explosions or coalescences of
binary neutron stars.  Our approach uses the weak-scale hidden sector models in
which the Higgs sector of the standard model is extended to include a gauge
singlet that only interacts with the Higgs particle.  These particles would be
produced either during the implosion of the red supergiant star core or at the
aftermath of a neutron star binary merger.  The whole energetics and timescales
of the relativistic blast wave, the fireball, are reproduced.
\end{abstract}


\narrowtext \twocolumn ]


The discovery of the afterglow \cite{afterglow} associated with some of the
gamma--ray bursts (GRBs) and isotropy of the emitted radiation both support the
view that the GRBs occur at cosmological distances with a redshift of the order
of one.  In this sense, GRBs offer a new observational tool for probing the early  universe.

There have been several suggestions concerning the generation mechanisms as
well as the distributions of photons in the core of the astrophysical object,
i.  e., the central engine of the GRBs\cite{suggestions}.  Among these, for
example, the resonant production of gamma rays during the collision of two
neutron stars is one possible mechanism \cite{sasha}.  Pertinent to a typical
neutron star, the core region of the progenitor has a characteristic radius of
$R_{0}\sim 10 \mbox{km}$ with roughly constant temperature $k_{B}T_{0}\sim 50
\mbox{MeV}$ and matter density $\rho_{0}\sim 10^{14} \mbox{g}/\mbox{cm}^{3}$.
The daily GRBs require a burst duration of $1~{\rm ms} \simlt \tau \simlt
0.1$~s~ with a total energy release of $\sim 10^{53} {\rm erg}$.

Given the extensive parameters of the progenitor and the observed power
spectrum, it is known that the gamma photons are trapped in the core region.
Then one is to think about other possible agents to transfer the correct amount
energy in the given time interval through the baryonic load.  One possible
alternative is neutrinos \cite{neutrino}; however, the mixing of the flavour
neutrinos with the sterile one is strongly suppressed in such matter densities,
and thus, the oscillation picture runs into difficulties \cite{amol-alexei}.
Another alternative mechanism would come through the axions; however, the
transferred power diminishes if the breaking scale of the Peccei--Quinn
symmetry gets higher \cite{axion1}.  In fact, the required axion mass cannot be
reproduced in known axion models at all \cite{axion2}.

In this letter we work out a different scenario for transporting the energy
outside the GRBs central engine.  The basic agent of the process is a CP--even,
presumably light, scalar particle, $S$, which has no baryonic charges.  Such
singlets have been proposed to take into account the nonobservation of the
standard model (SM) Higgs particle by increasing its invisible decay rate
\cite{singlet}.  For clarity of the discussion, we write down the effective
Lagrangian describing the interactions between this SM--singlet and the photons
as

\bea \label{lagran} - L_{int}\ni \frac{1}{2} m_{s}^{2} S^{2}+ \lambda_{S}
S^{4}+\lambda_{\gamma} S^{2} A_{\mu}A^{\mu} \eea

where $A_{\mu}$ is the photon, and $m_s$ and $\lambda_S$ designate the mass and
the quartic coupling of the singlet, respectively.  The original model
\cite{durmus} has an unbroken U(1) symmetry associated to the complex nature of
this singlet.  However, for the purpose of this work the global phase of the
singlet is not important so we take it real.  In the framework of the hidden
Higgs sector models the couplings above take the form

\begin{eqnarray} \lambda_S \approx \lambda_S^0+\kappa^2~,\ \ \ \lambda_\gamma
\approx \kappa \left(\frac{\alpha^3}{\pi}\right)^{1/2} A_{\gamma}
\end{eqnarray}

where $\lambda_S^0$ is the bare quartic coupling of the theory, and $\kappa~
m_{h}$ stands for $h S S$ coupling \cite{durmus}.  In writing these expressions
we neglected invariant masses in a given channel compared to the Higgs boson
mass.  The vertex factor $A_{\gamma}$ for $h \gamma \gamma$ coupling is a
slowly varying function of $\sqrt{s}$, and it is of the order of one
\cite{gunion}.  In all computations below we will parametrize results in terms
of $\lambda_S$ and $\lambda_\gamma$ without going back to relations above.
However, one keeps in mind that numerically $\lambda_\gamma \sim
{\cal{O}}(10^{-4})~\kappa$.

\begin{figure*}[hbt] \centerline{
\psfig{figure=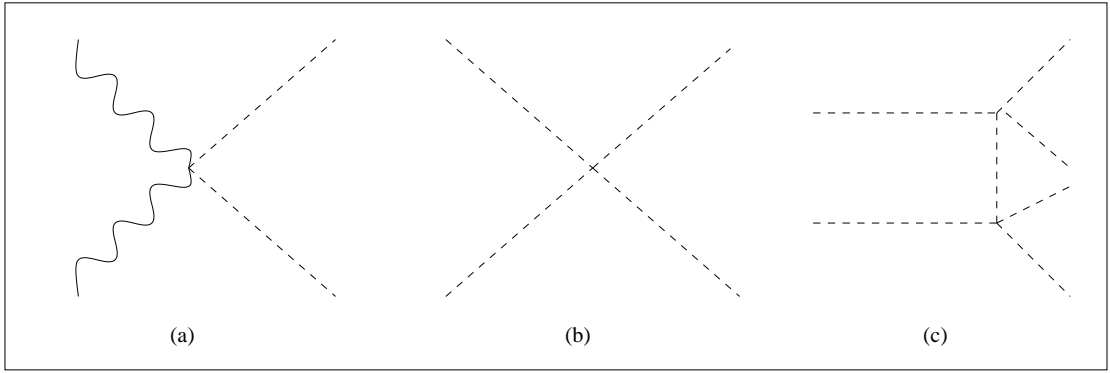,height=10cm,bbllx=1.0cm,bblly=9.cm,bburx=24.cm,bbury=22.cm}}
\caption{\footnotesize Relevant scattering processes following from the
interaction Lagrangian (\ref{lagran}) where the wavy lines represent the photon
and dashed lines the singlet.  The diagram (a) is responsible for conversion of
photons to singlets and vice versa.  On the other hand, the diagrams (b) and
(c) determine the mean free path of the singlets.}  \end{figure*}

The effective Lagrangian (\ref{lagran}) describes a real scalar field
interacting with itself and photons.  The scattering processes following from
this Lagrangian are shown in Fig.1.  Here the diagram (a) represents $\gamma
\gamma \leftrightarrow S S $ scattering with which the conversion of the
photons to singlets in the core of the GRBs burster occurs.  Furthermore,
conversion of singlets to photons outside (baryon--depleted region of) the GRBs
central engine happens with the same process in backward direction.  To have a
description of the energy transport through the strong baryonic load, it is
convenient to start with the conversion of the photons to the singlets.  The
relevant cross section reads as

\begin{eqnarray}\label{initcross} \sigma(\gamma \gamma \rightarrow S S ) =
\frac{\lambda_{\gamma}^{2}}{24 \pi s} \left(1 - \frac{4
m_{s}^{2}}{s}\right)^{1/2} \end{eqnarray}

where $\sqrt{s} \sim 100 {\rm MeV}$ is the total invariant mass of the
annihilating photons.  This process occurs in the core of the progenitor, and
the produced singlet pairs move out through the surrounding baryonic loading.
Before describing the journey of the singlets in the baryon load, it is
convenient to compute the rate of energy conversion from photons to singlets.
Denoting the four-momenta of the photons by $k_{1}=(\omega_{1}, \vec{k}_{1})$
and $k_{2}=(\omega_{2}, \vec{k}_{2})$, the total amount of energy converted to
singlet pairs per unit time per unit volume reads

\bea \label{qgamtos} Q=\int
\frac{d^{3}\vec{k}_{1}}{(2\pi)^{3}}\frac{d^{3}\vec{k}_{2}}{(2\pi)^{3}}n(\omega_{1})
n(\omega_{2}) (\omega_{1}+\omega_{2}) v_{rel} \sigma(\gamma\gamma\rightarrow S
S) \eea

where $n(\omega)$ is the equilibrium Bose population of the photons and
$v_{rel}=1 - \vec{k}_{1}\cdot\vec{k}_{2}/\omega_{1}\omega_{2}\equiv s/(2
\omega_1 \omega_2)$ is the relative velocity of the two annihilating photons.
After integrating $Q$ over the volume of the core region, the total luminosity
for photon--to--singlet conversion becomes

\begin{eqnarray} {\cal{L}}_{\gamma \gamma \rightarrow S S} \approx
\lambda_{\gamma}^{2} 10^{70} \left( \frac{T}{T_0} \right)^{5} \left(
\frac{R}{R_0} \right)^{3}~ {\rm erg} \cdot {\rm sec}^{-1} \end{eqnarray}

where ${\cal{O}}(m_s^{2}/(k_B T)^{2})$ terms are neglected in computing $Q$.
This is a good approximation in the burster core where temperature is high
enough.  Assuming that this will be the luminosity observed on Earth, a
comparison with the GRBs standard candle luminosity requires
$\lambda_{\gamma}\sim 10^{-8}$.  As will be discussed later, $\lambda_{\gamma}$
is a loop--induced quantity in the weak--scale hidden Higgs sector models so
that such a small number is naturally expected there.  Despite all these,
however, what is observed on Earth is ${\cal{L}}_{S S \rightarrow \gamma
\gamma}$, that is, one has to convert the singlets back to photons to simulate
the experimental conditions so that these naive bounds on $\lambda_{\gamma}$
may vary.

The singlets, after being pair--produced by photon--photon annihilations in the
core region, travel through the strong baryonic load towards the
baryon--depleted region outside the GRBs burster.  Since there is no
interaction with the baryons they do not feel the baryonic load at all, and
would move freely along radially outward trajectories were it not for their
self-interactions.  It is clear that the temperature of the host baryon
distribution does not effect the distribution and dynamics of the singlets as
they can never come to thermal equilibrium with the baryons.  In this sense,
even if the star is quite cold with a small fraction of ${\rm MeV}$ temperature
in the optically thin region, the singlets themselves can be quite energetic to
generate the $S S \rightarrow \gamma \gamma$ reaction.  Namely what singlets
take out of the star is the energy accumulated in the gamma photons and this
happens independent of the temperature and density distribution of the baryons.

The singlet self-interactions are depicted in diagrams (b) and (c) of Fig.1.
Both diagrams are generated by the singlet quartic coupling in the Lagrangian.
A close inspection of these diagrams reveal some important properties.  The $S
S \rightarrow S S$ scattering depicted in Fig.1(b) preserves the number of
singlets, is a contact interaction, and is kinematically operative for
$\sqrt{s}\geq 2 m_s$.  The $S S \rightarrow S S S S $ scattering in Fig.1(c),
on the other hand, doubles the number of singlets, is a long--range interaction
with range $\sim m_s^{-1}$, and is kinematically allowed when $\sqrt{s}\geq 4
m_s$.  The relevant cross sections are estimated to be

\begin{eqnarray} \label{cross} \sigma\left({\rm Fig.}1(b)\right) =
\frac{\lambda_{s}^{2}}{16 \pi s}> > \sigma \left({\rm Fig.}1(c)\right) \sim
{\cal{O}}\left( \frac{\lambda_{s}^{4}}{(2 \pi)^{4}s} \right) \end{eqnarray}

so that cross section for $S S \rightarrow S S S S$ scattering is second order
in $\lambda_s$ and receives further phase space suppressions.

The most important quantity describing the motion of the singlets towards the
GRBs baryon--depleted region is their mean free path.  Having no interactions
with the baryons, the singlet mean free path would be infinitely long were it
not for the singlet self-interactions depicted in diagrams (b) and (c) of
Fig.1.  The total mean free path obeys the relation

\begin{eqnarray} \ell^{-1}=\ell^{-1}_{(b)} + \ell^{-1}_{(c)} \end{eqnarray}

where subscripts refer to the diagrams in Fig.1.  Therefore, the total mean
path over which singlets move freely is described by the larger of the
individual contributions (for Fig.1(b) and Fig.1(c), respectively)

\begin{eqnarray} \label{paths} \ell^{-1}_{(b,c)}=\int
\frac{d^{3}\vec{p_t}}{(2\pi)^{3}} n(E_t) v_{rel}\left(1 - \frac{4
m_s^2}{s}\right)^{1/2} \sigma \end{eqnarray}

where $n(E_t)$ is the equilibrium Bose population for the target singlet and
$v_{rel}\equiv s/(2 E E_t)$ is the relative velocity of the incident and target
singlets with respective four--momenta $p=(E, \vec{p})$ and $p_{t}=(E_t,
\vec{p_t})$.  Here we take the phase space density of singlets as a Bose
distribution ignoring the possibility of free streaming.  In any case the
resulting mean free path will be a conservative estimate of the actual one.

As Eq.(\ref{paths}) suggests clearly, larger the cross section smaller the
corresponding mean free path.  Using the expressions for the cross sections in
(\ref{cross}) one can make the rough estimate

\begin{eqnarray} \label{path} \ell\approx \ell_{(b)}\sim \left( \frac{E}{k_B
T_0}\right)~ \left(\frac{50 {\rm MeV}}{k_B T_0}\right) \left(
\frac{10^{-8}}{\lambda_s}\right)^{2} \sim 100 {\rm km} \end{eqnarray}

neglecting the terms ${\cal{O}} \left(m_s^2/k_B^{2} T_0^{2}\right)$.

This mean free path results solely from the self-interactions of the singlets,
that is, it is the singlets themselves which prevent their further flight.  In
particular, it is not the baryons that limit their motion so that it does not
matter if $\ell$ happens to fall inside or outside the baryon load region;
literally, singlets will stand still in a sphere of mean radius
${\cal{O}}(\ell)$ measured from the core of the progenitor.  To clarify this
point further one recalls, for instance, the neutrino propagation.  In that
case the mean free path is determined by the neutrino--baryon interactions, and
if it is outside the GRBs baryon load region neutrinos get out of the
astrophysical source otherwise they are trapped by the baryons and form a
thermal neutrino sphere.  Therefore, the formation of the singlet sphere
follows only from their self-interactions.

Due to their self-interactions in Fig.1(b) and (c) singlets will form a
thermalized cloud of particles whose number and energetics will change with a
chain of such scatterings.  As mentioned before, the $2\rightarrow 2$
scattering in Fig.1(b) preserves the number of singlets and plays an important
role in restricting the singlets to have the finite path (\ref{path}).  The
$2\rightarrow 4$ scattering in Fig.1(c), however, is a long--range interaction
and it modifies the number of singlets.  Due especially to its long--range
nature it is effective everywhere in the singlet sphere, and causes largely
separated singlet pairs to annihilate into four new singlets.  This
interaction, thus, increases the number of singlets and reduces the mean energy
per capita.  The resulting cloud of singlets will thermalize themselves with
their self-interactions with a temperature much lower than the burster core
temperature.  At any point inside the singlet sphere, there will be singlets
coming from every direction which is important in computing the energy
accumulation in a given region.  If singlets were moving along radially outward
trajectories there would be a strong geometrical suppression factor for the
energy deposition \cite{daretal}.

As mentioned above, because of $2\rightarrow 4$ process in Fig.1(c) total
number of singlets increases, and thus, average energy per singlet decreases.
Similar to the electromagnetic showers initiated by photons, a straightforward
computation of the total number of produced singlets as the mean energy per
singlet drops from $E_0$ to a critical energy $E_c$ via the $2\rightarrow 4$
scattering gives $N=N_0\left[1+ (E_0-E_c)/E_c\right]$ where $N_0$ is the
initial number of singlets given by the number of photons.  In the problem at
hand, $E_0\sim k_B T_0$ and $E_c \sim k_B T \sim 2 m_{s}$ where the latter
follows from the kinematic blocking of $2\rightarrow 4$ scatterings.  Number of
singlets per unit volume can be computed over a sphere of radius $\ell$:  $\rho
= 3 N/(4 \pi \ell^3)$.  Since the singlet cloud eventually thermalizes it is
convenient to use the usual Bose population for the number of singlets per unit
volume per unit momentum state with an appropriate scaling to reproduce the
singlet density $\rho$:

\begin{eqnarray} \rho = K \int \frac{d^{3} \vec{k}}{(2 \pi)^{3}} n(E)
\end{eqnarray}

where $n(E)$ is the Bose population of the singlets at temperature $T$, and $K$
is a constant for reproducing $\rho$.

The next thing one needs for computing the electromagnetic power generated by
the singlets is the $S S \rightarrow \gamma \gamma$ cross section

\begin{eqnarray} \label{fincross} \sigma(S S \rightarrow \gamma \gamma) =
\frac{\lambda_{\gamma}^{2}}{8 \pi s} \left(1 - \frac{4
m_{s}^{2}}{s}\right)^{-1/2} \end{eqnarray}

to be compared with $\gamma \gamma \rightarrow S S$ cross section in
(\ref{initcross}).  Then computation of the luminosity proceeds in exact
similarity with (\ref{qgamtos}) after replacing the cross section there by
(\ref{fincross}):

\begin{eqnarray} {\cal{L}}_{S S \rightarrow \gamma \gamma} \approx \left(
\frac{R_0}{\ell} \right)^{3} \left( \frac{T_0} {T} \right)^{3}
{\cal{L}}_{\gamma \gamma \rightarrow S S}, \end{eqnarray}

where a factor of $(T_0/T)^2$ follows from $(E_0/E_c)^2$.  ${\cal{L}}_{S S
\rightarrow \gamma \gamma}$ is computed over the volume of singlet sphere.
Thus, the larger the mean free path the higher the luminosity suppression
factor.  However, the higher the initial (core) temperature the higher the
luminosity emitted.  Since both effects scale with the same power it is clear
that no energy is lost in the global process, and thus the mechanism here
suggested is able to effectibly carry out the overall energy released in the
GRBs central engine as in the context of fireball models.  Once the energy is
out of the baryon loading, a relativistic blast wave of electron-positron pairs
and radiation is formed, i.  e., a fireball, which cleans away the burster
environment pushing out a rather matter free debris to produce the observed
burst when colliding with and external interstellar medium.  The no-time-delay
feature of this scenario makes it suitable to explain bursts with both rapid
risetime and prompt afterglows as observed in GRB990123.  The potential role of
this mechanism in triggering GRBs from supermassive star explosions is an issue
currently pursued.\cite{herman00}

To conclude, we have investigated the viability of the hidden Higgs sector
models of turning realizable the energy transport in GRBs in the context of
fireball scenarios\cite{Rees}.  As the discussions in the text show, in such
models transport of the energy from the core to outside is quite efficient, and
the resulting luminosity agrees with the astronomical observations.  The
particle physics scenarios with neutrinos and axions are not as efficient as
the present model due to the suppresion of the neutrino mixing angle and
smallness of the axions mass, respectively.  In the present model, conversion
processes are mediated by the Higgs particle.  On the other hand, transport of
the energy from the core to outside is done by the singlets having a rather
large mean free path compared to neutrinos.  In the present scenario singlets
are light enough to be pair-produced by the photon annihilations, and such a
light singlet does not contradict with the present collider data as it affects
the precision observables at two and higher loop levels\cite{singlet}.

The model employed here has two free parameters, the coupling constants
$\lambda_\gamma$ and $\lambda_S$, both hidding the explicit dependence on
 the singlet mass $m_{s}$ and Higgs mass $m_{h}$.  The first
parameter is constrained to be around $\lambda_\gamma \leq 10^{-8}$ by the
current GRBs BATSE observations, and its value will be measured at the $\gamma \gamma$ mode of the TESLA collider\cite{tesla}. As a final remark, the potential realization of this scenario in the astrophysical sources triggering GRBs would render it a viable pathway for testing some of the extensions of the standard model of particle physics introduced to account for the overall GRBs observational properties: energetics, timescales, spectra, etc., such as the one being suggested here.

\acknowledgements {We would like to thank  A. Dar, A. Kusenko, S. Nussinov and A. Yu. Smirnov for helpful suggestions and fruitful discussions on this work.}

 \end{document}